\documentclass[11pt]{article}
\usepackage[dvips]{graphicx}
\usepackage{amsmath}
\usepackage{amsxtra}
\usepackage{amssymb}
\begin{document}
\title { "Straightforward Derivation of the Schr{\"o}dinger Equation from Classical Mechanics and the Planck Postulate".}
\author{ A.Granik\thanks{Department of Physics, University of the
Pacific,Stockton,CA.95211} } \maketitle
\begin{abstract}
According to the widely accepted notion, the Schr{\"o}dinger equation (SE) is not derivable in principle. Contrary to this belief, we present here a straightforward derivation of SE. It is based on only two fundamentals of mechanics: the classical Hamilton-Jacobi equation(HJE) and the Planck postulate about the discrete transfer of energy at micro-scales. Our approach is drastically different from the other published derivations of SE which either employ an ad hoc underlying assumption about the probabilistic or the statistical nature of the micro-scale phenomena, or rely on the prior knowledge of SE and arrive at it by introducing a new postulate - neither present in classical mechanics nor following from experiments - with a suitable but physically unjustifiable choice of a key arbitrary constant.
\end{abstract}
\newpage
 \hspace*{4cm}
        \begin{minipage}{3in}
                {\small{"...with regard to the origin of the presented recipe the Schr{\"o}dinger equation we have to unequivocally say: there is no derivation and there cannot be any. To derive means to obtain it logically from other laws. The Schr{\"o}dinger equation has been guessed."}}\cite{LM}
        \end{minipage}

  \section{\large{Introduction}.}

Thirty years after the above categorical statement by L.I.Mandelshtam,
R. Feynman wrote \cite{f1}, "For a particle moving freely in space with no forces, no disturbances, the correct law of physics is
\begin{equation}
\label{1}
\int {H(x,x')\psi (x')dx' =  - \frac{{\hbar ^2 }}{{2m}}\frac{{d^2 }}{{dx^2 }}\psi }
\end{equation}
Where did we get that from? Nowhere. It is not possible to derive from anything you know. It came out of the mind of Schr{\"o}dinger, invented in his struggle to find an understanding of the experimental observation of the real world".
\footnote{in Eq.(\ref{1}) $\psi(x)$ is the wave function and $H(x',x)$ is the Hamiltonian  of the particle.}
	
In turn, the "inventor" himself, E. Schr{\"o}dinger, admitted  in the second of his seminal papers \cite{S1}:
"So far we have only briefly described this correspondence\footnote{namely, between the HJE of a mechanical problem and the "allied" wave equation. In Eq. \ref{2} and in what follows, S is the classical action of the particle.}  on its external analytical side by the transformation \cite{S2}
\begin{equation}
\label{2}
S = \frac{\hbar }{i}Ln\psi
\end{equation}
which is in itself unintelligible, and by equally incomprehensible transition from equating to zero a certain expression to the postulation that the space integral of the said expression shall be stationary".

After more than 40 years after Feynman's statement and more than 80 years after Schr{\"o}dinger's admission, there have been many approaches to derivation of the law (\ref{1}) from something "we know."              Referring only to the primary works these approaches have been based on \

{\hspace{-12pt}a) some ad hoc underlying assumptions, such as the probabilistic nature of quantum phenomena \cite{Nl}, or\

{\hspace{-12pt}b) an earlier analogous assumption about the statistical nature of quantum theory \cite {Ml}, or \

{\hspace{-12pt}c) a modification of the classical HJE by introducing into it the dissipative term \cite{AG}, or\

{\hspace{-12pt}d) the prior knowledge of the Schr{\"o}dinger equation to be arrived at by modifying the classical HJE via introduction of a new postulate (not present in classical mechanics, nor following from the experiments) and eventually, after a rather involved mathematics, equating - without a physical justification - a certain key arbitrary constant to the reduced Planck constant \cite{FM}.\

As to the Schr{\"o}dinger "unintelligible transformation"(\ref{2}), it has never been derived at all. It was used later by R. Feynman as an underlying postulate in his groundbreaking path integral formulation of quantum mechanics.\

In sharp contrast to the above works \cite{Nl}-\cite{FM}, in this paper\

{\hspace{-12pt}a) no assumptions about the character of the micro-scale phenomena are {\hspace{-12pt}made},\

{\hspace{-12pt}b) no new postulates into physics are introduced,\

{\hspace{-12pt}c)	no prior knowledge of the Schr{\"o}dinger  equation is assumed.\

 It is nonetheless shown that both fundamentals of quantum mechanics - the Schr{\"o}dinger equation (\ref{1}) and the critical relation (\ref{2})  -  are derivable in a straightforward (algorithmic) way by using only {\bf{two}} well-established and well-known $rudiments$ of physics:\

{\hspace{-12pt}1)	The classical relativistic HJE, which follows directly from Newton's second law of motion (see Appendix A\footnote{The numbers of equations in Appendix A have an extra letter A; for instance
Eq. (A11)}).\

{\hspace{-12pt}2)	The experimental fact of energy transfer on micro-scales in discrete quanta.

\section{\large{Massless Particle.}}

Let us begin from the relativistic Hamilton-Jacobi equation for a massless "particle" [see Appendix A, Eq.(\ref{A13})]
\begin{equation}
\label{3}
\left( {\frac{{\partial S}}{{\partial t}}} \right)^2 \; - \;c^2 \left( {\nabla S} \right)^2 \; = \;0
\end{equation}
This equation admits two solutions: the particle-like and the wave-like.
As shown below, these two solutions are not independent but related on a micro-scale via the discrete character of energy transfer. This turns out to be the "key" which enables us to derive the Schr{\"o}dinger equation from the above-mentioned rudiments.\

\subsection{\normalsize{Particle-Like Solution.}}

For simplicity sake, we consider the $1$-D case. A particle-like solution of (1) can be obtained by separating the variables:
\begin{equation}
\label{4}
S_p (x,t) =  - Et + f(x)
\end{equation}
where $E = constant$ is the energy of the massless "particle." Upon substitution of (\ref{4}) in (\ref{3}) and solving it with respect to $x$, we have
\begin{equation}
\label{5}
f = \frac{E}{c}x + {\rm constant}
\end{equation}
Without any loss of generality, we can set the above constant = 0. According to the definition of the action $S$, $df/dx=\partial S/\partial x=p_x\equiv p$  . On the other hand, from the energy-momentum relation for a massless particle, it follows
\begin{equation}
\label{6}
p=E/c
\end{equation}
Therefore a particle-like solution of (\ref{3}) is
\begin{equation}
\label{7}
S_p  =  - Et + px
\end{equation}

\subsection{\normalsize{Wave-Like Solution.}}

The wave-like solution of (\ref{3}) can be arrived at by transforming this relation into the wave equation. To this end, we differentiate (\ref{3}) with respect to $t$
\begin{equation}
\label{8}
\frac{{\partial S}}{{\partial t}}\frac{{\partial ^2 S}}{{\partial t^2 }} - c^2 \frac{{\partial S}}{{\partial x}}\frac{{\partial ^2 S}}{{\partial t\partial x}} = 0
\end{equation}
and then differentiate (\ref{3}) with respect to $x$
\begin{equation}
\label{9}
\frac{{\partial S}}{{\partial t}}\frac{{\partial ^2 S}}{{\partial t\partial x}} - c^2 \frac{{\partial S}}{{\partial x}}\frac{{\partial ^2 S}}{{\partial x^2 }} = 0
\end{equation}
Upon substitution of (\ref{9}) in (\ref{8}) we get
\begin{equation}
\label{10}
\frac{{\partial ^2 S}}{{\partial t^2 }} - c^4 \frac{{\partial ^2 S}}{{\partial x^2 }}\left( {\frac{{\partial S/\partial x}}{{\partial S/\partial t}}} \right)^2  = 0
\end{equation}
According to (\ref{3}), we have $(\partial S/\partial x)^2 /(\partial S/\partial t)^2  = 1/c^2 $  which on substitution in (8) yields the wave equation
\begin{equation}
\label{11}
\frac{{\partial ^2 S_w }}{{\partial t^2 }} - c^2 \frac{{\partial ^2 S_w }}{{\partial x^2 }} = 0
\end{equation}
For a monochromatic wave, a general solution to (\ref{11}) is a traveling wave of the frequency $\omega$  and the wave number $k$:
\begin{equation}
\label{12}
S_w  = ae^{i(kx - \omega t)} ,\quad a\; = \;{\rm constant}
\end{equation}\
Thus, by considering one classical equation (\ref{3}), we have obtained two solutions - relations (\ref{4}) and (\ref{12}) - which describe respectively two patterns of behavior of a massless particle:
\begin{itemize}
\item	the particle-like behavior, \

and
\item	the wave-like behavior.
\end{itemize}

As a result, we can say that a massless particle exhibits the so-called wave-particle duality feature: on the one hand it behaves like a massive particle and on the other hand it behaves like a wave.

\subsection{\normalsize{Derivation of the Schr{\"o}dinger Relation, Eq.(\ref{2}).}}

Now we are going to show that the two patterns of behavior of a massless particle are not independent. To find the relation between the particle-like, Eq. (\ref{4}), and the wave-like, Eq. (\ref{12}), solutions we differentiate the latter with respect to $t$ and obtain
\begin{equation}
\label{13}
\frac{{\partial S_w }}{{S_w \partial t}} \equiv \frac{\partial }{{\partial t}}\left( {\frac{{LnS_w }}{i}} \right) =  - \omega
\end{equation}
On the other hand,
\begin{equation}
\label{14}
\frac{{\partial S_p }}{{\partial t}} \equiv  - E
\end{equation}

To relate (\ref{14}) and (\ref{13}), we use Planck's hypothesis and its subsequent development by Einstein who has found that the electro-magnetic energy on a micro-scale is emitted and received in discrete quanta, that is,
\begin{equation}
\label{15}
E = \hbar \omega
\end{equation}
Thus at micro-scales the energy of a massless particle given by Eq. (\ref{14}) is equal to its expression given by (\ref{15}). Therefore by substituting (\ref{15}) in (\ref{14}), and using (\ref{13}) we get
\begin{equation}
\label{16}
\frac{{\partial S_p }}{{\partial t}} \equiv  - \hbar \omega  = \frac{\partial }{{\partial t}}(\frac{\hbar }{i}LnS_w )
\end{equation}
From (\ref{16}) follows
\begin{equation}
\label{17}
S_p  = \frac{\hbar }{i}Ln\psi
\end{equation}
where
\begin{equation}
\label{18}
\psi=AS_w,
\end{equation}
 $A$ is an arbitrary constant, and we call $\psi$ the wave function.\

Thus we have derived relation (\ref{17}) directly from classical mechanics (that is from "what we know") modified by the existence of the lower bound on the amount of energy transfer at micro-scales. This equation is the very same "unintelligible" transformation (\ref{2}) "invented" by Schr{\"o}dinger, and which was presumed impossible to derive "from anything we know".\

\subsection{\normalsize{Derivation of the Schr{\"o}dinger Equation, Eq. (\ref{1}).}}

By the definition, the particle's momentum $p_x=\partial S/\partial x$ . Upon substitution in this definition relations (\ref{17}) , (\ref{18}), and the wave solution (\ref{12})  we obtain the relation between the particle's momentum $p$ and the wave number $k$
\begin{equation}
\label{19}
\frac{{\partial S_p }}{{\partial x}} = p = \frac{\hbar }{i}\frac{1}{\psi }\frac{{\partial \psi }}{{\partial x}} = \hbar k
\end{equation}
This is the famous de Broglie relation for a massless particle, which he obtained with the help of rather contrived notions of internal and external oscillatory processes. If we identically rewrite Eq. (\ref{19}), we obtain the relation between the particle's momentum and the wave function in the following form
\begin{equation}
\label{20}
\frac{\hbar }{i}\frac{{\partial \psi }}{{\partial x}} = p\psi
\end{equation}
Quite similarly, we get from (\ref{14}), (\ref{17}), (\ref{18}) and (\ref{12})
\begin{equation}
\label{21}
 - E = \frac{{\partial S_p }}{{\partial t}} = \frac{1}{\psi }\frac{\hbar }{i}\frac{{\partial \psi }}{{\partial t}} \to i\hbar \frac{{\partial \psi }}{{\partial t}} = E\psi
\end{equation}
Thus for a massless particle its momentum and energy emerge as the eigenvalues of the operators $(\hbar/i)\partial/\partial x$  and $(i\hbar)\partial/\partial t$  acting on the wave function $\psi$.\

As the next step, we differentiate (\ref{20}) with respect to $x$ and (\ref{21}) with respect to $t$ and obtain respectively
\begin{equation}
\label{22}
\left( {\frac{\hbar }{i}\frac{\partial }{{\partial x}}} \right)\left( {\frac{\hbar }{i}\frac{\partial }{{\partial x}}} \right)\psi  = p\frac{\hbar }{i}\frac{{\partial \psi }}{{\partial x}} = p^2 \psi \quad or\quad  - \hbar ^2 \frac{{\partial ^2 \psi }}{{\partial x^2 }} = p^2 \psi
\end{equation}\

\begin{equation}
\label{23}
\left( {\hbar i\frac{\partial }{{\partial t}}} \right)\left( {\hbar i\frac{\partial }{{\partial t}}} \right)\psi  = E\left( {\hbar i\frac{{\partial \psi }}{{\partial t}}} \right) = E^2 \psi \quad or\quad  - \hbar ^2 \frac{{\partial ^2 \psi }}{{\partial t^2 }} = E^2 \psi
\end{equation}
We rewrite (\ref{22}) and (\ref{23}) one more time
\begin{eqnarray}
\label{24}
  - \hbar ^2 \frac{1}{\psi }\frac{{\partial ^2 }}{{\partial x^2 }}\psi  = p^2  \nonumber\\
  - \hbar ^2 \frac{1}{\psi }\frac{{\partial ^2 }}{{\partial t^2 }}\psi  = E^2
 \end{eqnarray}
On the other hand,
\begin{eqnarray}
\label{25}
 p^2  = (\frac{{\partial S}}{{\partial x}})^2  \nonumber\\
 E^2  = (\frac{{\partial S}}{{\partial t}})^2
 \end{eqnarray}\

	At the first glance, it seems that two different representations of $p^2$    and  $E^2$  in Eqs. (\ref{24}) and (\ref{25}) are not consistent.  However,this is not the case. In fact by using (\ref{17}) we have
\begin{eqnarray}\nonumber
  - \hbar ^2 \frac{1}{\psi }\frac{{\partial ^2 }}{{\partial x^2 }}\psi  = (\frac{{\partial S}}{{\partial x}})^2  =  - \hbar ^2 \frac{1}{{\psi ^2 }}(\frac{{\partial \psi }}{{\partial x}})^2 , \\
  - \hbar ^2 \frac{1}{\psi }\frac{{\partial ^2 }}{{\partial t^2 }}\psi  = (\frac{{\partial S}}{{\partial t}})^2  =  - \hbar ^2 \frac{1}{{\psi ^2 }}(\frac{{\partial \psi }}{{\partial t}})^2  \nonumber
 \end{eqnarray} 	
Rearranging these equations we obtain

\begin{eqnarray} \label{26}
\begin{gathered}
 \frac{{\psi \partial ^2 \psi /\partial x^2  - (\partial \psi /\partial x)^2 }}{{\psi ^2 }} \equiv \frac{\partial }{{\partial x}}(\frac{{\partial \psi /\partial x}}{\psi }) = \frac{{\partial ^2 }}{{\partial x^2 }}(Ln\psi ) = 0, \\
   \frac{{\partial ^2 \psi /\partial t^2  - (\dot \psi )^2 }}{{\psi ^2 }} = \frac{{\partial ^2 }}{{\partial t^2 }}(Ln\psi ) = 0\hfill \\
   \end{gathered}
 \end{eqnarray}
The solution of (\ref{26})
\begin{equation}
\label{27}
Ln\psi  = i(kx - \omega t)
\end{equation}
gives us the solution (\ref{12}) for the wave function of a free massless relativistic particle.\

The derived relations (\ref{22}) and (\ref{23}) allow us to obtain the Schr{\"o}dinger equation for a massless relativistic "particle", if we use relation $E =c p$, Eq.(\ref{6}),  which follows from the relativistic Hamilton-Jacobi equation (\ref{3}):
	 \begin{equation}
\label{28}
\begin{gathered}
  - \hbar ^2 \frac{{\partial ^2 }}{{\partial t^2 }}\psi  = E^2 \psi  = c^2 p^2 \psi  =  - c^2 \hbar ^2 \frac{{\partial ^2 }}{{\partial x^2 }}\psi \quad or \\
 \hbar ^2 \frac{{\partial ^2 }}{{\partial t^2 }}\psi  - c^2 \hbar ^2 \frac{{\partial ^2 }}{{\partial x^2 }}\psi  = 0 \hfill
 \end{gathered}
 \end{equation}
The last equation of (\ref{28}) is the Schr{\"o}dinger equation identical to the wave equation  (10), multiplied by $\hbar$, for a massless particle.

\section{\large{Massive Particle.}}

\subsection{\normalsize{Derivation of the 3-D Schr{\"o}dinger Equation.}}

 Let us consider a free relativistic particle of the rest mass $m_0$. The relativistic Hamilton -Jacobi equation for a free massive particle is\footnote{see Appendix A, Eq.(A11)}
\begin{equation}
\label{29}
(\frac{{\partial S_p }}{{\partial t}})^2  - c^2 (grad\;S_p )^2  = m_0 ^2 c^4 \end{equation}
A solution of (\ref{29}) is obtained by the separation of variables
\begin{equation}
\label{30}
S_p=-Et+\vec{p}\cdot\vec{r}
\end{equation}
Inserting (\ref{30}) in (\ref{29}) we get the special-relativistic energy-momentum formula
\begin{equation}
\label{31}
E^2  = p^2 c^2  + m_0 c^4
\end{equation}\

 	According to the Lemma $1$ in  Appendix B\footnote{Equation (B3)}, the substitution identical to Eq.(\ref{17}) for a massless particle
	\begin{equation}
\label{32}
S_p  = ALn\psi
\end{equation}
transforms equation (\ref{29}) into the homogeneous second order nonlinear PDF
\begin{equation}
\label{33}
 A^2 (\frac{{\partial \psi }}
{{\partial t}})^2  - c^2 A^2 (grad\;\psi )^2  = m_0 ^2 c^4 \psi ^2
\end{equation}
where constant $A$ to be determined. Furthermore, according to corollary to Lemma $2$ \footnote{Equation (B21) in Appendix B} equation (\ref{33}) is equivalent to the following wave equation:
\begin{equation}
\label{34}
A^2 \frac{{\partial ^2 \psi }}
{{\partial t^2 }} - A^2 c^2 (\frac{{\partial ^2 \psi }}
{{\partial x^2 }} + \frac{{\partial ^2 \psi }}
{{\partial y^2 }} + \frac{{\partial ^2 \psi }}
{{\partial z^2 }}) - m_0 ^2 c^4 \psi  = 0
\end{equation}
whose general solution is a traveling monochromatic wave
\begin{equation}
\label{35}
\psi=const\times exp(i\vec{k}\cdot\vec{r}-i\omega t),~~~   \vec{k}(k_x,k_y,k_z),~~~ \vec{r}(x,y,z)
\end{equation}
The respective dispersion relation $\omega=\omega(k)$ is given by Eq.(B15) in Appendix B:
\begin{equation}
\label{35a}
\omega  = \sqrt {c^2 k^2  - (m_0 c^2 /A)^2 }
\end{equation}

Thus, even a massive particle exhibits the wave-particle duality, formally expressed by the existence of two solutions: particle-like (\ref{30}) corresponding to the Hamilton-Jacobi equation (\ref{29}) and wave-like (\ref{35}) corresponding to the wave equation (\ref{34}).\

To find the value of constant A we substitute (\ref{32}) in the definition of the particle's energy (\ref{14}), use the solution (\ref{35}), and add the postulate that at micro-scales a massive particle's energy is transferred in "chunks" according to  Eq.(\ref{15}), that is
\begin{equation}
\label{36}
 - E =  - \hbar \omega  = \frac{{\partial S_p }}
{{\partial t}} = A\frac{1}
{\psi }\frac{{\partial \psi }}
{{\partial t}} =  - Ai\omega
\end{equation}
As a result we obtain
	 \begin{equation}
\label{37}
A = \frac{\hbar }{i}
\end{equation}
which upon substitution in (\ref{32}) yields for a free massive particle exactly the same transformation ("invented" by Schr{\"o}dinger) as we obtained in section $2$, Eq.(\ref{17}), for a massless particle.\

	Finally, by using (\ref{37}) in the wave equation (\ref{34}) we obtain  the Schr{\"o}dinger equation for a free relativistic particle
	 \begin{equation}
\label{38}
\hbar ^2 \frac{{\partial ^2 \psi }}
{{\partial t^2 }} - \hbar ^2 c^2 (\frac{{\partial ^2 \psi }}
{{\partial x^2 }} + \frac{{\partial ^2 \psi }}
{{\partial y^2 }} + \frac{{\partial ^2 \psi }}
{{\partial z^2 }}) + m_0 ^2 c^4 \psi  = 0
\end{equation}
From (\ref{37}) and (\ref{32}) also follows that relations (\ref{24})  derived for a free massless particle hold true for a free massive particle.
\subsection{\normalsize{Derivation of the De Broglie Relation.}}

Now  we will derive the de Broglie relation, Eq.(\ref{19}), for a massive particle.  By (\ref{30}) and (\ref{32}) we have
\begin{equation}
\label{39}
\nabla S_p  = \mathbf{p} = \frac{\hbar }
{i}\frac{grad\psi}
{\psi }
\end{equation}
Upon substitution of  (\ref{35})  in (\ref{39})  we obtain the de Broglie relation
\begin{equation}
\label{40}
\mathbf{p}=\hbar\mathbf{k}
\end{equation}
for a massive relativistic particle, this time rigorously derived from classical mechanics supplemented by the requirement of  the discrete character of energy transfer. This is in contradistinction to the method used by de Broglie who employed rather artificial concepts of the internal and external oscillations associated with a particle.\

Let  us return to the dispersion relation (\ref{35a}). By inserting (\ref{35a}) in (\ref{35}) we get
\begin{equation}
\label{41}
\omega  = \sqrt {c^2 k^2  + (\frac{{m_0 c^2 }}
{\hbar })^2 }
\end{equation}
From (\ref{41}) we find  the phase  and group  velocities corresponding to  the wave properties of a relativistic "particle" at micro-scales
\begin{equation}
\label{42}
\begin{gathered}
  v_{ph}  = \frac{\omega }
{k} = c\sqrt {1 + (\frac{{m_0 c}}
{{\hbar k}})^2 }  > c,\quad  \hfill \\
  \mathbf{v}_{gr}  = \frac{{d\omega }}
{{d\mathbf{k}}} = \frac{c^2 \mathbf{k}}
{\sqrt {c^2 k^2  + (m_0 c^2/\hbar)^2}}
\equiv \frac{\hbar \mathbf{k}/m_0}
{\sqrt {c^2 k^2  + (m_0 c^2/\hbar)^2}}, \hfill \\
  |\mathbf{v}_{gr} | = \frac{c}
{\sqrt {c^2 k^2  + (m_0 c^2/\hbar)^2}} < c \hfill \\
\end{gathered}
\end{equation}

On the other hand, the velocity of a relativistic particle is
$$\mathbf{v} = \frac{\mathbf{p}/m_0 }
{\sqrt {1 + (p/m_0 c)^2 } } = \frac{\hbar \mathbf{k}/m_0 }
{\sqrt {1 + (\hbar k/m_0 c)^2} } $$	
where we use (\ref{40}). Thus in this case the group velocity of the wave coincides with the particle velocity of the wave-particle complex.

\subsection{\normalsize{The Non-Relativistic Schr{\"o}dinger Equation.}}

To obtain the Schr{\"o}dinger equation for a free non-relativistic particle we eliminate the rest-energy  from (\ref{38}) by representing $\psi$ as follows
	 	\begin{equation}
\label{43}
\psi  = \psi _0 (x,y,z)e^{ - im_0 c^2 t/\hbar }
\end{equation}
Upon substitution of (\ref{43}) in (\ref{38}) we get
\begin{equation}
\label{44}
\hbar ^2 \frac{{\partial ^2 }}
{{\partial t^2 }}\psi _0  - 2i\hbar m_0 c^2 \frac{{\partial \psi _0 }}
{{\partial t}} - c^2 \hbar ^2 (\frac{{\partial ^2 }}
{{\partial x^2 }} + \frac{{\partial ^2 }}
{{\partial x^2 }} + \frac{{\partial ^2 }}
{{\partial x^2 }})\psi _0  = 0
\end{equation}
We divide both sides of (\ref{44}) by $c^2$ and take the non-relativistic limit $c\rightarrow\infty$:
\begin{equation}
\label{45}
\begin{gathered}
  \mathop {Lim}\limits_{c \to \infty } \left[ {\frac{{\hbar ^2 }}
{{c^2 }}\frac{{\partial ^2 }}
{{\partial t^2 }}\psi _0  - 2i\hbar m_0 c^2 \frac{{\partial \psi _0 }}
{{\partial t}} - \hbar ^2 \left( {\frac{{\partial ^2 }}
{{\partial x^2 }} + \frac{{\partial ^2 }}
{{\partial x^2 }} + \frac{{\partial ^2 }}
{{\partial x^2 }}} \right)\psi _0 } \right]\; =  \hfill \\
   =  - 2i\hbar m_0 \frac{{\partial \psi _0 }}
{{\partial t}} - \hbar ^2 \left( {\frac{{\partial ^2 }}
{{\partial x^2 }} + \frac{{\partial ^2 }}
{{\partial x^2 }} + \frac{{\partial ^2 }}
{{\partial x^2 }}} \right)\psi _0 =0 \hfill \\
\end{gathered}
\end{equation}
which results in
	\begin{equation}
\label{46}
i\hbar \frac{{\partial \psi _0 }}
{{\partial t}} + \frac{{\hbar ^2 }}
{{2m_0 }}\left( {\frac{{\partial ^2 }}
{{\partial x^2 }} + \frac{{\partial ^2 }}
{{\partial x^2 }} + \frac{{\partial ^2 }}
{{\partial x^2 }}} \right)\psi _0  = 0
\end{equation}
Equation (\ref{46}) is the non-relativistic Schr{\"o}dinger equation for a free particle of mass $m_0$.
\section{\large{Conclusions}.}

In this paper the Schr{\"o}dinger equation for a free particle has  been systematically and in a surprisingly elementary fashion derived directly from two fundamentals: the classical relativistic Hamilton-Jacobi equation and the Planck postulate.\

One of the corollaries of our derivation is  the crucial transformation (\ref{17}), lying at the heart of both the Schr{\"o}dinger approach and the Feynman path integral method \cite{f2}:
\begin{equation}
\label{47}
\psi  = const\times\exp (iS_p /h)
\end{equation}
All the major architects of quantum mechanics Schr{\"o}dinger \cite{f1}, Dirac \cite{PD}, and Feynman \cite{f2}  introduced (\ref{47}) as an ad hoc postulate without any physical justification. In our paper this crucial relation emerges as a natural consequence of the wave-particle duality. This implies that at micro-scales the wave function $\psi$ replaces the particle's classical action $S_p$, thus  reflecting the dual character of phenomena at such scales.\
One more important corollary is that the canonical representations of particle's momentum and energy as the eigenvalues of the partial differential operators  are also the consequences of the dual character of the phenomena at micro-scales.\\

{\bf{Acknowledgments}}. The author would like to thank V.Granik for illuminating discussions of the manuscript.

\section{\large{Appendix A}.}
\renewcommand{\theequation}{A\arabic{equation}}\setcounter{equation}{0}

\subsection{\normalsize{Derivation of Relativistic HJE from Newton's 2nd Law}}

Let us consider the second law of Newton for a relativistic particle in the potential field $\Phi(\mathbf{r}),~\mathbf{r}(x,y,z)$
\begin{equation}
\label{A1}
\frac{{d\mathbf{p}}}
{{dt}} =  - \frac{{\partial \Phi }}
{{\partial \mathbf{r}}}
\end{equation}
Here  the relativistic momentum {\bf{p}} of the particle is
\begin{equation}
\label{A2}
\mathbf{p}=\frac{m_0\mathbf{v}}{\sqrt{1-v^2/c^2}}.~~v^2=\mathbf{v}\cdot\mathbf{v}
\end{equation}
For the following  we express the particle's velocity $\mathbf{v}$ in terms of its momentum $\mathbf{p}$. We obtain from (\ref{A2}):
	 	\begin{equation}
\label{A3}
\mathbf{v} = \frac{{\mathbf{p}/m_0 }}
{{\sqrt {1 + p^2 /m_0 ^2 c^2 } }},~~p^2=\mathbf{p}\cdot\mathbf{p}
\end{equation}

Now we consider the momentum $\mathbf{p}$ as a function of time $t$ and spatial coordinates $x,y,z$. This means replacement of the Lagrange description of motion by the Euler description. In the latter instead of moving with  the particle, as in the former,  a fixed observer watches the different path ,which the particle can follow at different moments of time. As a result we rewrite (\ref{A1}) as follows
\begin{equation}
\label{A4}
\begin{gathered}
  \frac{{d\mathbf{p}}}
{{dt}} = \frac{{\partial \mathbf{p}}}
{{\partial t}} + (\mathbf{v}\cdot grad)\mathbf{p} = \frac{{\partial \mathbf{p}}}
{{\partial t}} + \frac{1}
{{m_0 \sqrt {1 + p^2 /m_0 ^2 c^2 } }}(\mathbf{p}\cdot grad)\mathbf{p} \equiv  \hfill \\
  \frac{{\partial \mathbf{p}}}
{{\partial t}} + \frac{1}
{2m_0}\frac{1}
{{ \sqrt {1 + p^2 /m_0 ^2 c^2 } }}[grad(p^2 ) - \mathbf{p} \times curl\mathbf{p}] =  - grad\Phi  \hfill \\
\end{gathered}
\end{equation}
where we use the vector identity
$$(\mathbf{p}\cdot grad)\mathbf{p} \equiv \frac{1}
{2}[grad(p^2 ) - \mathbf{p} \times curl\mathbf{p}]$$
Since
\begin{equation}
\label{A5}
grad(\sqrt {1 + p^2 /m_0 ^2 c^2 } ) = \frac{1}
{2}\frac{{grad(p^2 )}}
{{m_0 ^2 c^2 \sqrt {1 + p^2 /m_0 ^2 c^2 } }}
\end{equation} 	
equation (\ref{A4}) becomes
\begin{equation}
\label{A6}
\frac{{\partial \mathbf{p}}}
{{\partial t}} + m_0 c^2 grad\sqrt {1 + p^2 /m_0 ^2 c^2 }  - \frac{1}
{2m_0}\frac{\mathbf{p} \times curl\mathbf{p}}
{ \sqrt {1 + p^2 /m_0 ^2 c^2 } } =  - grad\Phi
\end{equation}

If we apply operation curl to both sides of (\ref{A6}) we get
\begin{equation}
\label{A7}
\frac{\partial }
{{\partial t}}(curl\mathbf{p}) - \frac{1}
{{2m_0 }}curl(\frac{{\mathbf{p} \times curl\mathbf{p}}}
{{\sqrt {1 + p^2 /m_0 ^2 c^2 } }}) = 0
\end{equation}
	Equation (\ref{A7}) is identically satisfied, if
	$$curl\mathbf{p} = 0$$ 	
This means that
\begin{equation}
\label{A8}
\mathbf{p} = gradF
\end{equation}
where $F(r,t)$ is a scalar function. As the next step we substitute  (\ref{A8}) in equation (\ref{A4}) and obtain the following equation
\begin{equation}
\label{A9}
grad(\frac{\partial }
{{\partial t}}F + m_0 c^2 \sqrt {1 + (gradF)^2 /m_0 ^2 c^2 }  + \Phi ) = 0
\end{equation}
Its solution is
\begin{equation}
\label{A10}
\frac{\partial }
{{\partial t}}F + m_0 c^2 \sqrt {1 + (gradF)^2 /m_0 ^2 c^2 }  + \Phi  = f(t)
\end{equation}
where $f(t)$ is an arbitrary function of $t$.

We introduce a new function
\begin{equation}
\label{A11}
S = F(\mathbf{r},t) - \int {f(t)dt}
\end{equation}
Upon substitution of (\ref{A11}) in (\ref{A10}) we obtain the relativistic Hamilton-Jacobi equation for a free particle of mass $m_0$
\begin{equation}
\label{A12}
(\frac{{\partial S}}
{{\partial t}} + \Phi )^2  - c^2 (gradS)^2  - m_0 ^2 c^4  = 0
\end{equation}
For a free massless particle $m_0=0$, and equation (\ref{A12}) yields
\begin{equation}
\label{A13}
\left( {\frac{{\partial S}}
{{\partial t}}} \right)^2  - c^2 (gradS)^2  = 0
\end{equation}\newpage
\section{\large{Appendix B}.}
\renewcommand{\theequation}{B\arabic{equation}}\setcounter{equation}{0}
\subsection{\normalsize{Auxiliary Mathematical Lemmas.}}
{\bf{Lemma 1}}

A nonlinear partial differential equation of the $m$-th order with $n$ arguments $x_j$, $j=1,2,3,...,n$
\begin{eqnarray}
\label{B1}
\sum\limits_{k = 1}^n {a_k \frac{{\partial y}}
{{\partial x_k }}}  + \sum\limits_{j,k = 1}^n {a_{jk} \frac{{\partial y}}
{{\partial x_j }}\frac{{\partial y}}
{{\partial x_k }}}  + ...\nonumber\\
+\sum_{j,k,...,s=1}a_{jk...s}\overbrace{{\frac{{\partial y }}{{\partial x_j }}\frac{{\partial y}}{{\partial x_k }}}...\frac{{\partial\psi}}{{\partial x_s }}}^{m}+b=0
\end{eqnarray}
is transformed into the homogeneous nonlinear equation of the same order
\begin{equation}
\begin{gathered}
\label{B2}
\sum\limits_{k = 1}^n {a_k A\psi ^{m - 1} \frac{{\partial \psi }}
{{\partial x_k }}}  + \sum\limits_{j,k = 1}^n {a_{jk} A^2 \psi ^{m - 2} \frac{{\partial \psi }}
{{\partial x_j }}\frac{{\partial \psi }}
{{\partial x_k }}}  + ...\\ +\sum_{j,k,...,s=1}a_{jk...s}A^m\overbrace{{\frac{{\partial\psi }}{{\partial x_j }}\frac{{\partial\psi}}{{\partial x_k }}}...\frac{{\partial\psi}}{{\partial x_s }}}^{m}+b\psi^m=0
\end{gathered}
\end{equation}
by  the following transformation of the function $y$:
\begin{equation}
\label{B3}
y = ALn\psi
\end{equation}
where $A$ is a constant and the new function $\psi  = \psi (x_1 ,x_2 ,...,x_n)$\\

\hspace{-12pt}{\bf{Proof}}

Let us consider a transformation
\begin{equation}
\label{B4}
y = y(\psi )
\end{equation}
where $\psi  = \psi (x_1 ,x_2 ,...,x_n)$. We require this new
function $y(\psi)$ to be such that upon substitution of (\ref{B4}) in (\ref{B1}) it would transform it into equation (\ref{B2}). Upon substitution of (\ref{B4}) in (\ref{B1}) we get
\begin{equation}
\label{B5}
\begin{gathered}
\sum_{k = 1}^n {a_k\frac{dy}{d\psi} \frac{{\partial\psi}}
{{\partial x_k }}}  + \sum_{j,k = 1}^n {a_{jk}(\frac{dy}{d\psi})^2 \frac{\partial\psi}
{\partial x_j }\frac{\partial\psi}{\partial x_k }}  + ...\\
+\sum_{j,k,...,s=1}a_{jk...s}(\frac{dy}{d\psi})^m\overbrace{{\frac{{\partial\psi }}{{\partial x_j }}\frac{{\partial\psi}}{{\partial x_k }}}...\frac{{\partial\psi}}{{\partial x_s }}}^{m}+b=0
\end{gathered}
\end{equation}

	Assuming that $dy/d\psi  \ne 0$  we divide both sides of (\ref{B5}) by $(dy/d\psi)^m$
\begin{equation}
\label{B6}
\begin{gathered}
\sum\limits_{k = 1}^n {a_k \frac{1}
{{(dy/d\psi )^{m - 1} }}\frac{{\partial \psi }}
{{\partial x_k }}}  + \sum\limits_{j,k = 1}^n {a_{jk} \frac{1}
{{(dy/d\psi )^{m - 2} }}\frac{{\partial \psi }}
{{\partial x_j }}\frac{{\partial \psi }}
{{\partial x_k }}}  + ...\\
+\sum_{j,k,...,s=1}a_{jk...s}\overbrace{{\frac{{\partial\psi}}{{\partial x_j }}\frac{{\partial\psi}}{{\partial x_k }}}...\frac{{\partial\psi}}{{\partial x_s }}}^{m}+\frac{b}{(dy/d\psi)^m}=0
\end{gathered}
\end{equation}
Equation (B6) is transformed into equation (B2) if
\begin{equation}
\label{B7}
\frac{1}{dy/d\psi } = \frac{\psi }{A}
\end{equation}
or equivalently
\begin{equation}
\label{B8}
\frac{dy}{d\psi } = \frac{A}{\psi}
\end{equation}
where $A$ is a constant.

	Solving differential equation (\ref{B8}) we obtained
\begin{equation}
\label{B9}
y = ALn\psi
\end{equation}
The constant of integration has been included into the new function $\psi$.
Conversely, upon substitution of  (\ref{B9}) in (\ref{B1}) we obtain equation (\ref{B2}).\\

\hspace{-12pt}{\bf{Lemma $2$}}

A homogeneous nonlinear  partial differential equation of the $m-th$ order ($m=2,3,...$) with $n$ arguments $x_j (j = 1,2,...,n)$
 \begin{equation}\label{B10}
 \begin{gathered}
 \sum\limits_{k = 1}^n {a_k A\psi ^{m - 1} \frac{{\partial \psi }}
{{\partial x_k }}}  + \sum\limits_{j,k = 1}^n {a_{jk} A^2 \psi ^{m - 2} \frac{{\partial \psi }}
{{\partial x_j }}\frac{{\partial \psi }}
{{\partial x_k }}}  + ...\\ +\sum_{j,k,...,s=1}a_{jk...s}A^m\overbrace{{\frac{{\partial\psi }}{{\partial x_j }}\frac{{\partial\psi}}{{\partial x_k }}}...\frac{{\partial\psi}}{{\partial x_s }}}^{m}+b\psi^m=0
\end{gathered}
  \end{equation}
and constant coefficients $b$, $a_j ,a_{jk,} ...,a_{\mathop {\underline {jk...s} }\limits_m }$($j,k,...=1,2,...n $)has  the oscillatory solution $\exp (i\sum\limits_{j=1}^n\alpha _j x_j ),\;i \equiv \sqrt { - 1} $ [with the dispersion relation  $f(\alpha_j,a_j,a_{jk},...,b)=0$]\\

\hspace{-12pt}{\bf{Proof}}
	
We restrict our proof to a case  important in physical applications, when the order of homogeneity is $m=2$ and the number of arguments $n=4$,  with $3$ spatial arguments $x_{1,} x_2 ,x_3 $, and the temporal argument $x_4 = t$. In this case Eq.(\ref{B10}) written in the Cartesian coordinates is
\begin{equation}
\label{B11}
a_{jk} A^2 \frac{{\partial \psi }}{{\partial x_j }}\frac{{\partial \psi }}
{{\partial x_k }} + b\psi ^2  = 0
\end{equation}
where the coefficients $a_{jk} ,b$ are real-valued and we use the convention of summation over the repeated indices.\
	
Upon substitution of
\begin{equation}
\label{B12}
\psi  = const\times\exp (i\alpha _l x_l ),\quad l = 1,2,3,4
\end{equation}
in equation (\ref{B11}) we obtain the dispersion equation  $f(\alpha_j)=0$ where $\alpha_k~(k=1,2,3)$ have the physical meaning of the components $k_x,k_y,k_z$ of the wave vector $\vec{k}$ and $\alpha_4$ has the physical meaning of the wave frequency $\omega$
\begin{equation}
\label{B13}
\begin{gathered}
  A^2 a_{44} \omega ^2  + \omega \;A^2 [k_x (a_{14}  + a_{41} ) + k_y (a_{24}  + a_{42} ) + k_z (a_{34}  + a_{43} )] +  \hfill \\
   + A^2 [(a_{12}  + a_{21} )k_x k_y  + (a_{13}  + a_{31} )k_x k_z  + (a_{23}  + a_{32} )k_y k_z ] +  \hfill \\
   + A^2 (a_{11} k_x ^2  + a_{22} k_y ^2  + a_{33} k_z ^2 ) - b = 0 \hfill
\end{gathered}
\end{equation}
Solution of this equation gives us the dispersion relation $\omega  = f\left( k \right)$ for the monochromatic wave described by  solution (\ref{B12}) of equation (\ref{B11}).

For the particular case corresponding to the Hamilton-Jacobi equation for a free particle we have $b =  - m_0 ^2 c^4$ ($c$ is the speed of light) and
\begin{equation}
\label{B14}
a_{jk}  = \left\{ \begin{gathered}
   - c^2 ,\quad j = k \ne 4, \hfill \\
  1,\quad \quad j = 4, \hfill \\
  0,\quad \quad j \ne k \hfill \\
\end{gathered}  \right.
\end{equation}
The respective solution of (\ref{B13}) is then
\begin{equation}
\label{B15}
\omega  = \sqrt {c^2 k^2  - (m_0 c^2 /A)^2 }
\end{equation}

Conversely, given function (\ref{B12}) and the dispersion relation (\ref{B13}) we obtain
\begin{equation}
\label{B16}
\frac{{\partial \psi }}
{{\partial x_j }}\frac{{\partial \psi }}
{{\partial x_k }} = \alpha _j \alpha _k \psi ^2
\end{equation}
As the next step, we multiply (\ref{B16}) by $A^2 a_{jk} $, sum the result over all the values of $j$  and $k$ ( in our particular case $j,k =1,2,3,4$) and add the term $b\psi^2$:
\begin{equation}
\label{B17}
A^2 a_{jk} \frac{{\partial \psi }}
{{\partial x_j }}\frac{{\partial \psi }}
{{\partial x_k }} + b\psi ^2  \equiv (A^2 a_{jk} \alpha _j \alpha _k  + b)\psi ^2
\end{equation}

Since by (\ref{B13})  the expression in parentheses of (\ref{B17}) is equal to zero, this proves that function (\ref{B3}) is the solution of the original differential equation (\ref{B11}).\\

\hspace{-12pt}{\bf{Corollary}}.

Equation (\ref{B11}) is equivalent to the following second order linear partial differential equation
\begin{equation}
\label{B18}
A^2 a_{jk} \frac{{\partial ^2 \psi }}
{{\partial x_j \partial x_k }} + b\psi  = 0
\end{equation}
with the same constant coefficients as in (\ref{B11}).\\

\hspace{-12pt}{\bf{Proof}}.

We identically rewrite (\ref{B18}) as follows
\begin{equation}
\label{B19}
\begin{gathered}
  A^2 a_{jk} \frac{{\partial \psi }}
{{\partial x_j }}\frac{{\partial \psi }}
{{\partial x_k }} + b\psi ^2  \equiv \psi (A^2 a_{jk} \frac{{\partial ^2 \psi }}
{{\partial x_j \partial x_k }} + b\psi ) +  \hfill \\
   + A^2 \psi ^2 a_{jk} [\frac{1}
{{\psi ^2 }}(\frac{{\partial \psi }}
{{\partial x_j }}\frac{{\partial \psi }}
{{\partial x_k }} - \frac{{\partial ^2 \psi }}
{{\partial x_j \partial x_k }})] = 0 \hfill \\
\end{gathered}
\end{equation}
The expression in the last brackets is
\begin{equation}\label{B20}
\frac{1}
{{\psi ^2 }}(\frac{{\partial \psi }}
{{\partial x_j }}\frac{{\partial \psi }}
{{\partial x_k }} - \frac{{\partial ^2 \psi }}
{{\partial x_j \partial x_k }}) \equiv \frac{\partial }
{{\partial x_j }}(\frac{\partial }
{{\partial x_k }}Ln\psi ) = \frac{{\partial ^2 }}
{{\partial x_j \partial x_k }}Ln\psi
\end{equation}

On the other hand, the original equation (\ref{B11}) has a solution (\ref{B12}), that is $\psi  = const\exp (i\alpha _l x_l )$. Using this solution in (\ref{B20}) we find that
$$\frac{{\partial ^2 }}
{{\partial x_j \partial x_k }}Ln\psi  = 0$$	 	
Inserting this back in (\ref{B19}) we obtain the wave equation (\ref{B18}). For the special case given by (\ref{B14}) equation (\ref{B11}) is equivalent to the following equation
\begin{equation}
\label{B21}
A^2 \frac{{\partial ^2 \psi }}
{{\partial t^2 }} - A^2 c^2 \left( {\frac{{\partial ^2 \psi }}
{{\partial x^2 }} + \frac{{\partial ^2 \psi }}
{{\partial y^2 }} + \frac{{\partial ^2 \psi }}
{{\partial z^2 }}} \right) - m_0 ^2 c^4 \psi  = 0
\end{equation}
\end{document}